\documentclass[amstex,aps,preprint,nofootinbib]{revtex4}%
\usepackage{graphicx}
\usepackage{color}
\usepackage{bm}
\usepackage{braket}
\usepackage{longtable}
\usepackage{amsmath}
\usepackage{amsfonts}
\usepackage{amssymb}
\usepackage{float}
\usepackage{epstopdf} 
\begin{document}
\title{ Weak values and quantum properties}
\author{A. Matzkin}
\affiliation{Laboratoire de Physique Th\'{e}orique et Mod\'{e}lisation (CNRS Unit\'{e}
8089), Universit\'{e} de Cergy-Pontoise, 95302 Cergy-Pontoise cedex, France}
%\author{A}
%\affiliation{AA}

\begin{abstract}
We investigate in this work the meaning of weak values through the prism of property ascription in quantum systems. Indeed, the weak measurements framework contains only ingredients of the standard quantum formalism, and as such weak measurements are from a technical  point of view uncontroversial. However attempting to describe properties of quantum systems through weak values -- the output of weak measurements -- goes beyond the usual interpretation of quantum mechanics, that relies on eigenvalues. We first recall the usual form of property ascription, based on the eigenstate-eigenvalue link and the existence of ``elements of reality''. We then describe against this backdrop the different meanings that have been given to weak values. We finally argue that weak values can be related to a specific form of property ascription, weaker than the eigenvalues case but still relevant to a partial description of a quantum system.

\end{abstract}
\maketitle

\section{Introduction}

Weak values and weak measurements were introduced by Aharonov, Albert and
Vaidman 30 years ago \cite{AAV} as a tool to understand the properties of
quantum systems at intermediate times between preparation and a final state
obtained by measuring a chosen observable. Initially applied to elucidate
apparently paradoxical behavior in quantum systems, such as the
three-box-paradox \cite{AV-3box}, weak measurements have become increasingly
popular in the last 10 years, in part due to several experiments that were
able to observe weak values \cite{expt}, and also due to promising
technological applications, in particular regarding the amplification of weak
signals \cite{amplification}. Weak measurements have also been claimed to play
a role in reformulating quantum theory \cite{aharonov-botero,hoffmann2015}.

Although the theoretical framework of weak measurements (WM) and weak values
(WV) only involves ingredients of standard quantum mechanics, WM and WV were
criticized \cite{leggett,peres} very soon after their inception, and the
criticism has persisted to this day. We will leave aside the substantial
fraction of the criticism that is rooted in misunderstandings or erroneous
readings of the formalism. Instead the bulk of the arguments, while
recognizing the formalism is correct, denies that WM have anything to say
concerning the properties of quantum systems at intermediate times. Actually,
in the first published criticism of the WM framework, Leggett had already asserted that in his view,
weak measurements could not be qualified to be measurements at all \cite{leggett}.

The reason WM and WV have remained controversial is that although the
formalism is unambiguous, in the sense that all the ingredients come from
standard quantum mechanics, the latter remains silent on many interpretative
issues, and in particular has no provision to account for the properties of a
system without making a projective measurement -- but doing so renders the very
question adressed by WM (to understand the properties at an intermediate time)
meaningless, since a projective measurement radically modifies the
system evolution. Hence the idea at the basis of WM: induce a weak coupling
between a system observable and a quantum pointer, a coupling so weak that the
evolving state vector is minimally modified and that the probability of
obtaining a given outcome when making the final measurement is not modified
relative to the no coupling situation. The weak value is precisely the number
characterizing the motion of the quantum pointer due to the weak interaction.

The question is then whether this weak value (or rather, its real part, since,
as we will recall below, WV are complex) can be taken as a generalized form of
eigenvalue as advocated in the original paper \cite{AAV}, or are instead
meaningless or arbitrary numbers or at any rate useless to describe the
properties of a quantum system at an intermediate time. The rationale for the
latter position is that WV can lie outside the eigenvalue spectrum, so that
for example the weak value of a projector can be negative. Should this be
taken as yet another strange quantum feature, or does it mean there is an
irremediable flaw when attempting to attribute a value to a quantum property
through weak values? What is at stake here is not only the status of weak
values, but more fundamentally the relevance of the results obtained within
the weak measurements framework in order to understand the physical nature of
quantum systems. Indeed, WM open a new observational window into the quantum
world, allowing to acquire information on a system without substantially modifying
its evolution. It is crucial to assess the nature of this information, viz.
whether it is related to the properties that are weakly measured.

In the present manuscript, we investigate these questions by reexaming how a
property value is ascribed to a quantum system. We start by discussing the
eigenstate-eigenvalue link, which is the basis of property ascription in
quantum systems.\ We introduce the notion of pre-selection and post-selection
and examine how the eigenstate-eigenvalue link ascribes properties in such
circumstances, i.e. state preparation (pre-selection) followed by an
intermediate projective measurement and finally post-selection (filtering of a
particular outcome of a final projective measurement of a different
observable). We then introduce the Weak Measurements framework (Sec.
\ref{sec-WM}) and give a few properties of weak values that are important in
the present context. Sec. \ref{interpretation} critically examines the
different meanings that have been given to weak values. Indeed, by
construction, property ascription for weak values cannot rely on the
eigenstate-eigenvalue link, and WV have therefore been related to other
features (such as conditional averages over statistical ensembles or response
functions to a small perturbation).\ We will nevertheless argue (Sec
\ref{wvwv}) that there is room to relate weak values to quantum properties but
in a very specific, elusive manner, in a much weaker way than what is provided
by the eigenstate-eigenvalue link. We then expose our view on the meaning of
weak values (Sec. \ref{meaning}) and finally present our Conclusions in
Sec.\ \ref{concl}.

\section{Properties in quantum systems \label{prop}}

\subsection{The eigenstate-eigenvalue link\label{eel}}

The standard approach to quantum mechanics is to ascribes a property to a
quantum system when the system is in an eigenstate of a given observable. If a
given property is represented by an observable $A$ with eigenstates
\ $\left\vert a_{k}\right\rangle $ and eigenvalues $a_{k}$, ie
\begin{equation}
A\left\vert a_{k}\right\rangle =a_{k}\left\vert a_{k}\right\rangle
\end{equation}
(assumed here discrete and non-degenerate), then if the system is in a state
$\left\vert \psi\right\rangle $, that can generally be represented as
\begin{equation}
\left\vert \psi\right\rangle =\sum_{k}c_{k}\left\vert a_{k}\right\rangle ,
\end{equation}
the value of the property represented by $A$ is not defined, unless
$\left\vert \psi\right\rangle $ is an eigenstate of $A$ (in which case all the
$c_{k}$ vanish except one). 

The fact that a definite value cannot be ascribed
to an observable in an arbitary state was already quite clearly stated in
Dirac's early textbook (see Secs 9 and 10 of Ref.\ \cite{dirac}): ``\emph{The
expression that an observable `has a particular value' for a particular state
is permissible in quantum mechanics in the special case when a measurement of
the observable is certain to lead to the particular value, so that the state
is an eigenstate of the observable}''. Otherwise Dirac writes that ``\emph{a
disturbance involved in the act of measurement causes a jump in the state}'' of
the system (\cite{dirac}, p. 36).\ This approach, a cornerstone of the orthodox
interpretation, is often known as the eigenstate-eigenvalue link (see Ref. \cite{eev-history} for a historical account of the term).

In his textbook \emph{Quantum Mechanics}, a masterly exposition of the
orthodox approach, Bohm explains in addition that a given property value only
appears when the system is actually measured, after it has interacted with a
measuring apparatus \cite{Bohm}. The physical underlying model is due to von
Neumann \cite{VN}. In von Neumann's impulsive measurement model, the quantum states of a
measuring pointer are explicitly introduced. Suppose that initially (at
$t=t_{i}$) the system is prepared into the state $\left\vert \psi
(t_{i})\right\rangle $.\ Let $\left\vert \varphi(t_{i})\right\rangle $
designate the initial state of the quantum pointer. The total initial quantum
state is the product state%
\begin{equation}
\left\vert \Psi(t_{i})\right\rangle =\left\vert \psi(t_{i})\right\rangle
\left\vert \varphi(t_{i})\right\rangle . \label{inis}%
\end{equation}
We assume the pointer state is initially compactly localized around some
position $x_{0}$; we will use the notation $\left\langle x\right\vert
\left.  \varphi_{x_{0}}\right\rangle =\varphi_{x_{0}}(x)$. Assume further that the system
and the pointer interact during a brief time interval $\tau$. Let $A$ be the
measured system observable. The interaction between the system and the quantum
pointer is given by the coupling Hamiltonian%
\begin{equation}
H_{int}=g(t)AP. \label{Hint}%
\end{equation}
$g(t)$ is a smooth function that vanishes for times $t\leqslant t_{i}$ or
$t\geqslant t_{i}+\tau$ and such that $g\equiv\int_{t_{i}}^{t_{i}+\tau}g(t)dt$
appears as the effective coupling constant. If $g$ is large we can neglect the
self Hamiltonians of the system and of the pointer and consider that the
evolution during the short time interval $\tau$ is solely driven by
$H_{int\text{ . }}$This leads to%
\begin{align}
\left\vert \Psi(t_{i}+\tau)\right\rangle  &  =e^{-igAP/\hbar}\left\vert
\psi(t_{i})\right\rangle \left\vert \varphi_{x_{0}}\right\rangle \\
&  =\sum_{k}\left\vert a_{k}\right\rangle \left\langle a_{k}\right\vert
\left.  \psi(t_{i})\right\rangle e^{-iga_{k}P/\hbar}\left\vert \varphi_{x_{0}%
}\right\rangle \\
&  =\sum_{k}\left\langle a_{k}\right\vert \left.  \psi(t_{i})\right\rangle
\left\vert a_{k}\right\rangle \left\vert \varphi_{x_{0}-ga_{k}}\right\rangle ,
\label{vn}%
\end{align}
where we have used in the last line the properties of the translation operator.

Eq. (\ref{vn}) associates each pointer state $\left\vert \varphi_{x_{0}%
-ga_{k}}\right\rangle $ (shifted relative to the initial pointer state by a
distance proportional to the eigenvalue $a_{k}$) with the corresponding
eigenstate $\left\vert a_{k}\right\rangle $. At this post-interaction stage,
we still have an entangled state: the interaction Hamiltonian (\ref{Hint})
drives the system to the observable eigenstates, but not yet to a definite
eigenvalue.\ In some sense, the system has acquired the property (the one
represented by the measured observable) relative to the pointer, but not yet
its value. A definite value only appears when the linear superposition
(\ref{vn}) is replaced by a single term corresponding to the observed value.
There is no consensus on the origin or nature of this collapse (that can be
taken as apparent or fundamentally real, depending on the specific
interpretation \cite{collapse}), though it has to do with some irreversible amplification
that takes place at the macroscopic scale when the pointer is measured. The
overall process described by von Neumann's model is known by the rather
syncretic term of \textquotedblleft projective measurement\textquotedblright.

The eigenstate-eigenvalue link calls therefore for an interaction between
\ the system and the pointer with a large coupling constant (large meaning
that the shift is larger than the spatial width of the initial state
$\varphi_{x_{0}}(x)$) and a collapse to a final pointer state unambiguously
correlated with an eigenstate of the measured observable.

\subsection{Element of reality\label{er}}

The eigenstate eigenvalue link is intimately related to the notion of
``elements of reality'', as introduced by Einstein, Podolsky and Rosen (EPR) \cite{EPR}:
``\emph{If, without in any way disturbing a system, we can predict with
certainty (i. e. , with probability equal to unity) the value of a physical
quantity, then there exists an element of physical reality corresponding to
this physical quantity}''. 

Indeed, it can be noticed that Eq. (\ref{vn}) is an
entangled state, not unlike an EPR pair. Measuring the pointer shift to be
$ga_{k_{1}}$ immediately correlates with the system eigenstate being
$\left\vert a_{k_{1}}\right\rangle $. This can be checked by repeating the
same measurement (with a second, identical pointer). We know with certainty
that the pointer will be shifted by the quantity $ga_{k_{1}}$.
Hence, because of the correlation encapsulated in the entangled state, after the measurement the system is with certainty in the state
$\left\vert a_{k_{1}}\right\rangle$. This implies that the corresponding eigenvalue is an ``element of reality'', and property ascription follows from there  since
we know with certainty the system property and its value. 

The relation between
the eigenstate-eigenvalue link and an element of reality was already noted by
Redhead \cite{redhead}, who coins this relation the \textquotedblleft
Eigenvector rule\textquotedblright\ (see Ch.\ 3 of Ref. \cite{redhead}).
Redhead also notes that the \textquotedblleft no disturbance\textquotedblright%
\ condition in EPR's definition of \textquotedblleft elements of
reality\textquotedblright\ is unnecessary (and even potentially confusing) as
far as the Eigenvector Rule is concerned. Hence we can state that when the eigenvalue-eigenstate link holds the
corresponding property can be ascribed to a quantum system.\ This property is
then an element of reality.

\subsection{Expectation values}

As is well known, there is no consensus as to whether the state vector
provides a description (complete or incomplete) of an individual system, or
describes instead an ensemble of similarly prepared systems, although the
standard view has increasingly tilted toward the statistical approach to state
vectors (see eg Ch.\ 9 of Ref.\ \cite{ballentine}). Expectation values however are never assumed to refer to properties of a single system.
An expectation value is instead obtained when the system is prepared in state $\left\vert
\psi(t_{i})\right\rangle $ and the measurement of the property represented by
\thinspace$A$, as described by the von Neumann model given above, is repeated
several times, with random outcomes $a_{k}$ obtained with probability
$p_{k}=\left\vert \left\langle a_{k}\right\vert \left.  \psi(t_{i}%
)\right\rangle \right\vert ^{2},$ leading to the standard expression%
\begin{equation}
\left\langle A\right\rangle _{\left\vert \psi(t_{i})\right\rangle
}=\left\langle \psi(t_{i})\right\vert A\left\vert \psi(t_{i})\right\rangle
=\sum p_{k}a_{k}. \label{ev}%
\end{equation}
In each run, the system ends up in the eigenstate $\left\vert a_{k}%
\right\rangle $ -- the system has the property given by the corresponding
eigenvalue -- but the expectation value is obviously not an \textquotedblleft
element of reality\textquotedblright.

\subsection{Counterfactuals}

It is intuitively tempting to go beyond the eigenstate-eigenvalue link and attempt to
ascribe properties to a quantum system as the system evolves from its initial
state to the final state obtained as the result of a projective measurement.
This can only be done by counterfactual reasoning. Indeed, ascribing a value
to a property would involve performing a projective measurement at some
intermediate time, but doing so would modify the original experimental
arrangement and affect the system evolution dramatically (the system may not
even reach the original final state). 
%A well-known simple instance is
%Wheeler's delayed choice experiment. Mach-Zehnder output ports etc.

Counterfactual definiteness conflicts with quantum mechanics on the general
ground \cite{vaidman-counter} that it leads to ascribe to quantum systems
joint properties that can never be simultaneously measured. This point was
made early on by Bohr, in particular in his reply \cite{Bohr-EPR} to EPR \cite{EPR}. Bohr
writes there that \textquotedblleft\emph{we have in each experimental
arrangement suited for the study of proper quantum phenomena not merely to do
with an ignorance of the value of certain physical quantities, but with the
impossibility of defining these quantities in an unambiguous way}%
\textquotedblright, which can be seen as vindicating the eigenstate-eigenvalue
link in order to ascribe properties. Bohr further pointed out that
countefactual reasoning usually leads to paradoxes.

\subsection{Properties in pre and post-selected systems:\ the ABL\ rule
\label{sec-abl}}

Pre-selected and post-selected systems are systems for which not only is the
initial state prepared in a known state (this is the pre-selected state) but
also the final state is fixed (this state is known as the post-selected state).\ In practice post-selection is performed by
filtering the outcome of the final projective measurement. This is
particularly useful when starting from a preselected state $\left\vert
\psi(t_{i})\right\rangle $, an intermediate standard projective measurement of
some observable $A$ is made before a final measurement of a different
observable $B$ takes place. The ABL rule \cite{ABL} states how to compute
probabilities for the outcomes $a_{k}$ of $A$ when the system has been
preselected in state $\left\vert \psi(t_{i})\right\rangle $ and will finally
be found in the post-selected eigenstate $\left\vert b_{f}\right\rangle $ of
$B$. The probability of obtaining $a_{n}$ in the intermediate measurement is
given by%
\begin{equation}
P(a_{n}|\psi(t_{i}),b_{f})=\frac{\left\vert \left\langle b_{f}\right\vert
\left.  a_{n}\right\rangle \left\langle a_{n}\right\vert \left.  \psi
(t_{i})\right\rangle \right\vert }{\sum_{k}\left\vert \left\langle
b_{f}\right\vert \left.  a_{k}\right\rangle \left\langle a_{k}\right\vert
\left.  \psi(t_{i})\right\rangle \right\vert ^{2}}. \label{abl}%
\end{equation}

The ABL rule is a standard quantum mechanical result that follows from the
Bayes rule and the Born rule. It illustrates that ascribing properties to a
quantum system is a delicate task. Consider indeed a particle that is allowed
to take 3 paths, eg a spin-1 charged particle in a Stern-Gerlach like setup\footnote{This implementation of the three-box paradox \cite{AV-3box} has been described in details elsewhere \cite{matzkin-pan,duprey2017}}.
Let the pre and post-selected states be given by%
\begin{align}
\left\vert \psi(t_{i})\right\rangle  &  =\left(  \left\vert \psi
_{1}\right\rangle +\left\vert \psi_{2}\right\rangle +\left\vert \psi
_{3}\right\rangle \right)  /\sqrt{3}\\
\left\vert b_{f}\right\rangle  &  =\left(  \left\vert \psi_{1}\right\rangle
-\left\vert \psi_{2}\right\rangle +\left\vert \psi_{3}\right\rangle \right)
/\sqrt{3}, \label{pitch}
\end{align}
where $\left\vert \psi_{j}\right\rangle $ denotes the state vector on path $j
$. We want to compute the probability of finding the particle on path 1
(conditioned on obtaining the final state $\left\vert b_{f}\right\rangle $).
The result depends however on how the measurement is implemented (and hence
how the observable is defined). If a projective measurement is made on
\emph{each path} then the
eigenstates $\left\vert a_{k}\right\rangle $ that can be obtained are
$\left\vert \psi_{1}\right\rangle $, $\left\vert \psi_{2}\right\rangle $ and
$\left\vert \psi_{3}\right\rangle $ and Eq. (\ref{abl}) yields $P(a_{1})=1/3$.
If instead we measure path-1 vs. non-path-1 (the latter being measured for instance by connecting paths 2 and 3 together and placing a measurement apparatus at that point, see Fig. 1 of \cite{matzkin-pan}), then the eigenstates $\left\vert a_{k}\right\rangle $
that can be obtained are now $\left\vert \psi_{1}\right\rangle $ and
$\left\vert \psi_{2}\right\rangle +\left\vert \psi_{3}\right\rangle $ and Eq.
(\ref{abl}) leads to $P(a_{1})=1$; this implies that $P(a_{2+3})=0$, a
straightforward consequence of the fact that the eigenstate $\left\vert
\psi_{2}\right\rangle +\left\vert \psi_{3}\right\rangle $ obtained at the
intermediate time is orthogonal to the post-selected state $\left\vert
b_{f}\right\rangle $ given by Eq. (\ref{pitch}). This means that in this situation, we are certain to find the particle on path 1.

Note that according to our analysis in Sec. \ref{prop}, the fact that the
system will be on path 1 with unit probability is an unambiguous property of the system, an element of reality. This is not
an innocuous remark, because as is well-known \cite{AV-3box}, we can repeat
exactly the same argument for a path 3 vs non-path 3 measurement (we now have a measurement apparatus on path 3 and another apparatus at a point where paths 1 and 2 are connected). That is if we compute the probability
$P(a_{3})$ to find the particle on path 3 on a path 3 vs. non-path 3
measurement we find $P(a_{3})=1$ and
$P(a_{1+2})=0$. We are thus certain to find the system on path 3. This
apparent paradox has triggered vivid discussions on
counterfactuals in pre and post-selected systems
\cite{vaidman-counter,kastner-counter,kirkpatrick,mohrhoff,vaidman-shmp}. We will just note here that
both properties following from $P(a_{1})=1$ \ and $P(a_{3})=1$ are
well-defined, but each in its own configuration, involving different
measurements and experimental arrangements. For instance when measuring path 1 vs. non-path 1, $\left\vert \psi_{3}\right\rangle $ is not an eigenstate of the corresponding measurement, and no
value can be ascribed to the property \textquotedblleft the particle is on
path 3\textquotedblright (contrarily to a path 3 vs. non-path 3 measurement). A paradox
only appears if counterfactuals are employed, and value assignment is made
without reference to the eigenstate-eigenvalue link. Conversely, embracing the eigenstate-eigenvalue link dispels the paradox but evades the question concerning the value of the path projectors at intermediate times. This is the difficulty that weak measurements aim to bypass.

\section{Weak Measurements\label{sec-WM}}

\subsection{Weak measurement protocol}\label{protocol}

Weak measurements \cite{AAV} deal with extracting information about  a
given property, represented by an observable $A$, as the system evolves from a
prepared initial state towards the final eigenstate obtained after measuring a
different observable $B$. The context is identical to the one exposed above
concerning the ABL rule, Sec. \ref{sec-abl}: the system is prepared in the
pre-selected state $\left\vert \psi(t_{i})\right\rangle $, a weak measurement
of $A$ takes place and finally $B$ is measured and outcomes corresponding to
the post-selected state $\left\vert b_{f}\right\rangle $ of $B$ are filtered.
The difference is that $A$ is not measured through a standard projective
measurement that would bring the system to one of the eigenstates. Instead, a
very weak interaction is established between the system and a quantum pointer,
so as to leave the system state \textquotedblleft essentially
undisturbed\textquotedblright, meaning that the perturbation is so small that
the post-selection probabilities are not affected by the weak interaction.

Let us therefore represent the initial system-pointer state as in Eq.
(\ref{inis}) by
\begin{equation}
\left\vert \Psi(t_{i})\right\rangle =\left\vert \psi(t_{i})\right\rangle
\left\vert \varphi(t_{i})\right\rangle . \label{inis2}%
\end{equation}
We will take the system-pointer interaction to be given again by
$H_{int}=g(t)AP$ as in Eq. (\ref{Hint}).\ Let us assume that the interaction
takes place in a time window $[t_{w}-\tau/2,t_{w}+\tau/2],$ i.e. $t_{w}$ is
the average interaction time and $\tau$ the duration. If $\tau$ is small
relative to the system evolution timescale, the interaction can be simply
taken to take place precisely at $t_{w}$ (for a proof of this
\textquotedblleft midpoint rule\textquotedblright, see Ref. \cite{A2012PRL}).
As in von Neumann's impulsive measurement scheme (see below Eq. (\ref{Hint}),
$g\equiv\int_{t_{w}-\tau/2}^{t_{w}+\tau/2}g(t)dt$ appears as the effective
coupling constant, but we now require $g$ to be very small. Finally, we will
allow for the system to evolve from $t_{i}$ to $t_{w}$ and denote
$U(t_{w},t_{i})$ the corresponding unitary operator, but disregard instead the
self-evolution of the pointer. After the interaction $(t>t_{w}+\tau/2)$ the
initial uncoupled state (\ref{inis2}) becomes :
\begin{align}
\left\vert \Psi(t)\right\rangle  &  =U(t,t_{w})e^{-igAP}U(t_{w},t_{i}%
)\left\vert \psi(t_{i})\right\rangle \left\vert \varphi(t_{i})\right\rangle
\label{corrsl20}\\
&  =U(t,t_{w})e^{-igAP}\left\vert \psi(t_{w})\right\rangle \left\vert
\varphi(t_{i})\right\rangle \label{corrsl2}\\
&  =U(t,t_{w})\sum_{k}e^{-iga_{k}P}\left\langle a_{k}\right\vert \left.
\psi(t_{w})\right\rangle \left\vert a_{k}\right\rangle \left\vert
\varphi(t_{i})\right\rangle . \label{corrs}%
\end{align}

At time $t_{f}$ the system undergoes a standard projective measurement of the
observable $B$. Filtering the results of this projective measurement by
keeping only projections to the postselected state $\left\vert b_{f}%
\right\rangle $ yields
\begin{equation}
\left\vert \varphi(t_{f})\right\rangle =\sum_{k}\left[  \left\langle
b_{f}(t_{w})\right\vert \left.  a_{k}\right\rangle \left\langle a_{k}%
\right\vert \left.  \psi(t_{w})\right\rangle \right]  e^{-iga_{k}P}\left\vert
\varphi(t_{i})\right\rangle , \label{finalps}%
\end{equation}
where we have used $\left\langle b_{f}(t_{w})\right\vert =\left\langle
b_{f}(t_{f})\right\vert U(t_{f},t_{w})$. $\varphi(x,t_{f})$ is then given by a
superposition of shifted initial states We now use the fact that the coupling
$g$ is small, so that $e^{-iga_{k}P}\approx1-iga_{k}P$ holds for each $k$. Eq.
(\ref{finalps}) takes the form%
\begin{align}
\left\vert \varphi(t_{f})\right\rangle  &  =\left\langle b_{f}(t_{w}%
)\right\vert \left.  \psi(t_{w})\right\rangle \left(  1-igP\frac{\left\langle
b_{f}(t_{w})\right\vert A\left\vert \psi(t_{w})\right\rangle }{\left\langle
b_{f}(t_{w})\right\vert \left.  \psi(t_{w})\right\rangle }\right)  \left\vert
\varphi(t_{i})\right\rangle \label{za1}\\
&  =\left\langle b_{f}(t_{w})\right\vert \left.  \psi(t_{w})\right\rangle
\exp\left(  -igA_{f}^{w}P\right)  \left\vert \varphi(t_{i})\right\rangle
\label{finwv}%
\end{align}
where%
\begin{equation}
A_{f}^{w}=\frac{\left\langle b_{f}(t_{w})\right\vert A\left\vert \psi
(t_{w})\right\rangle }{\left\langle b_{f}(t_{w})\right\vert \left.  \psi
(t_{w})\right\rangle } \label{wvt}%
\end{equation}
is the weak value of the observable $A$ given pre and post-selected states
$\left\vert \psi\right\rangle $ and $\left\vert b_{f}\right\rangle $
respectively. We will drop the index $f$ and write $A^{w}$ instead whenever
the post-selection state is uniquely fixed and no confusions may arise.

Note that here we have not explicitly included the pointer coupled to $B$, which
is a standard von Neumann pointer described in Sec.\ \ref{eel}.\ Hence when
referring to a pointer in the remainder of the text, we will usually mean the
weakly coupled quantum pointer that registers the weak measurement. This
pointer is a quantum system that will need to be measured in order to extract
the weak value.

\subsection{Weak values: properties}

The interpretative questions relative to the property of the system at the
intermediate time $t_{w}$
will be examined below in Sec. \ref{interpretation}.\ Here we note a few basic
properties of weak values that will be useful in our discussion below.

\subsubsection{Real part}

The first point to note is that in general the weak value is a complex
quantity. For an initially localized pointer state $\left\vert \varphi_{x_{0}%
}\right\rangle ,$ Eq. (\ref{finwv}) can be written as
\begin{equation}
\left\vert \varphi(t_{f})\right\rangle =\left\langle b_{f}(t_{w})\right\vert
\left.  \psi(t_{w})\right\rangle \exp\left(  g\operatorname{Im}A_{f}%
^{w}P\right)  \left\vert \varphi_{x_{0}-g\operatorname{Re}A^{w}}\right\rangle
. \label{finre}%
\end{equation}
The real part $\operatorname{Re}A^{w}$ induces a shift $\left\vert
\varphi_{x_{0}-g\operatorname{Re}A^{w}}\right\rangle $.\ This is similar to
the first step, Eq. (\ref{vn}), of the standard projective measurement, except
that here $g$ is small: the original and the shifted pointer states are almost
overlaping, so that extracting $\operatorname{Re}A^{w}$ cannot be done by
performing a single measurement of the pointer, contrary to the case of strong
$g$ which discriminates pointer states correlated with different eigenvalues
$a_{k}$.\ Note that if the pre or post-selected states $\left\vert \psi
(t_{w})\right\rangle $ or $\left\vert b_{f}(t_{w})\right\rangle $ is an
eigenstate of $A$, the weak value is real -- it is actually the corresponding
eigenvalue of $A.$ In the general case, $\operatorname{Re}A^{w}$ is different
from the eigenvalues and can lie outside the spectrum of $A$. From Eq.
(\ref{wvt}) it is straightforward to obtain%
\begin{equation}
\operatorname{Re}A^{w}=\frac{\left\langle \psi(t_{w})\right\vert \frac{1}%
{2}\left(  \Pi_{b_{f}(t_{w})}A+A\Pi_{b_{f}(t_{w})}\right)  \left\vert
\psi(t_{w})\right\rangle }{\left\langle \psi(t_{w})\right\vert \Pi
_{b_{f}(t_{w})}\left\vert \psi(t_{w})\right\rangle } \label{rea}%
\end{equation}
where $\Pi_{b_{f}(t_{w})}\equiv\left\vert b_{f}(t_{w})\right\rangle
\left\langle b_{f}(t_{w})\right\vert $ is the projector to the post-selected
state evolved backward in time to the time $t_{w}$ of interaction. Eq.
(\ref{rea}) has the form of a conditional expectation value when the system is
in state $\left\vert \psi(t_{w})\right\rangle $: the denominator is the
average of the projector $\Pi_{b_{f}(t_{w})}$ (i.e., the probability of
post-selection) while the numerator is the average of the symmetrized operator
$\Pi_{b_{f}(t_{w})}A+A\Pi_{b_{f}(t_{w})}$ (measurement of $A$ and projection
to $\left\vert b_{f}(t_{w})\right\rangle $).

Note that in the special case $\left\vert b_{f}(t_{w})\right\rangle =$
$\left\vert \psi(t_{w})\right\rangle $ (this happens in particular when there
is no self-evolution and the pre and postselected states are the same)
$\operatorname{Re}A^{w}=$ $\left\langle \psi(t_{w})\right\vert A\left\vert
\psi(t_{w})\right\rangle $ becomes a standard expectation value.

\subsubsection{Imaginary part}

The imaginary part can be put in the form%
\begin{equation}
\operatorname{Im}A^{w}=\frac{\left\langle \psi(t_{w})\right\vert \frac{1}%
{2i}\left(  \Pi_{b_{f}(t_{w})}A-A\Pi_{b_{f}(t_{w})}\right)  \left\vert
\psi(t_{w})\right\rangle }{\left\langle \psi(t_{w})\right\vert \Pi
_{b_{f}(t_{w})}\left\vert \psi(t_{w})\right\rangle }. \label{ima}%
\end{equation}
The numerator represents the average backaction of the measurement of $A$ on
the post-selection projector. This can be seen from the Liouville equation,
where the commutator $-i[\Pi_{b_{f}(t_{w})},A]$ appears as generating the
evolution of $\Pi_{b_{f}(t)}$ due to the interaction Hamiltonian Eq.
(\ref{Hint}) coupling $A$ to the quantum pointer. For the case $\left\vert
b_{f}(t_{w})\right\rangle =$ $\left\vert \psi(t_{w})\right\rangle ,$
$\operatorname{Im}A^{w}=0$.

\subsubsection{Expectation value}

The expectation value of $A$ in state $\left\vert \psi(t_{w})\right\rangle ,$
written in the standard form
\begin{equation}
\left\langle \psi(t_{w})\right\vert A\left\vert \psi(t_{w})\right\rangle
=\sum_{k}a_{k}p_{k}(a_{k}) \label{sav}%
\end{equation}
when $A$ is measured through a projective measurement, with $p_{k}%
(a_{k})\equiv\left\vert \left\langle a_{k}\right\vert \left.  \psi
(t_{w})\right\rangle \right\vert ^{2}$, can also be written as%
\begin{equation}
\left\langle \psi(t_{w})\right\vert A\left\vert \psi(t_{w})\right\rangle
=\sum_{k}\left\vert \left\langle b_{k}(t_{w})\right\vert \left.  \psi
(t_{w})\right\rangle \right\vert ^{2}\operatorname{Re}\frac{\left\langle
b_{k}(t_{w})\right\vert A\left\vert \psi(t_{w})\right\rangle }{\left\langle
b_{k}(t_{w})\right\vert \left.  \psi(t_{w})\right\rangle } \label{wav}%
\end{equation}
after some manipulations (see Eqs. (12)-(15) of \cite{duprey2018}), by which it can
also be seen that the weighted sum over the imaginary parts vanishes, so that
Eq. (\ref{wav}) can equivalently be written as
\begin{equation}
\left\langle \psi(t_{w})\right\vert A\left\vert \psi(t_{w})\right\rangle
=\sum_{k}A_{k}^{w}p_{k}(b_{k}), \label{wav2}%
\end{equation}
with $p_{k}(b_{k})\equiv\left\vert \left\langle b_{k}(t_{w})\right\vert
\left.  \psi(t_{w})\right\rangle \right\vert ^{2}$. Eqs. (\ref{wav}%
)-(\ref{wav2}) involve a projective measurement of $B$ and a weak measurement
of $A$.\ Relative to Eq. (\ref{sav}), the probabilities are now those of
obtaining a given post-selected state $\left\vert b_{k}\right\rangle $ while
the eigenvalues are replaced by the real part of the weak values associated
with the post-selected state $\left\vert b_{k}\right\rangle $.

\section{What do weak values stand for?\label{interpretation}}

\subsection{Preliminary remarks}

As mentioned in the Introduction, since its inception, weak values have
remained controversial, stirring much discussion. The fact that experimentally
the predictions of the weak measurements framework are verified is beyond
discussion. This is why the debate has centered on the meaning and
significance of the weak values. The viewpoint developed in this paper is to
frame this issue under the question: \textquotedblleft Is a weak value related
to a property of the system?\textquotedblright. To this end we recalled in
Sec.\ \ref{eel} the eigenstate-eigenvalue link, the basis of property
ascription in standard quantum mechanics. We have then seen in Sec.
\ref{sec-WM} that the weak value appears as a shift in the pointer state [Eq.
(\ref{finwv})], pretty much like an eigenvalue [Eq. (\ref{vn})]; the analogy
is also patent when comparing the expressions for the observable average
(\ref{sav}) and (\ref{wav2}) in terms of eigenvalues and weak values respectively. We have also seen however that the real and
imaginary parts of a weak value can be written in terms of conditional
expectations, Eqs. (\ref{rea}) and (\ref{ima}), making weak values look like an average. On the other hand, from its
definition, Eq. (\ref{wvt}) the weak value is seen to be the ratio of two
transition matrix elements, hence weak values are akin to amplitudes.

We will further analyze here the different meanings that the weak values can
take. An important point to keep in mind, obvious for practioners of weak
measurements but potentially confusing for others, is that previous to the the
measurement of $B$ the system state is undefined, as Eq. (\ref{finalps})
represents an entangled system-pointer state. After post-selection, at
$t=t_{f}$, the final system state $\left\vert b_{f}\right\rangle $ is an
eigenstate of $B,$ and according to the eigenstate-eigenvalue link, the system
has at that point acquired the property value $b_{f}$. The weak value also
becomes instantiated at $t=t_{f}$, although, as is clear from the definition
(\ref{wvt}), the weak value depends on the physical interaction that took
place at time $t_{w}$ (when the system interacted with the pointer). The weak
value is hence defined retroactively, as if the post-selected state had
propagated backwards in time. This does not call for any sort of
retrocausation (except if one endorses \cite{price2012} a time-symmetric
formulation of quantum mechanics, such as the Two State Vector Formalism
\cite{TSVF,TSQM}), but is a peculiar feature arising from quantum correlations
(see Secs. \ref{wvwv} and \ref{meaning}).

\subsection{Weak values and the eigenstate-eigenvalue link\label{wv-eel}}

By construction, weak measurements do not respect the eigenstate-eigenvalue
link. Indeed, the rationale is that the coupling between $A$ and the quantum
pointer should minimally disturb the system state, that is the coupling must
leave the post-selection probability $\left\vert \left\langle b_{k}%
(t_{w})\right\vert \left.  \psi(t_{w})\right\rangle \right\vert ^{2}%
=\left\vert \left\langle b_{k}(t_{f})\right\vert \left.  \psi(t_{f}%
)\right\rangle \right\vert ^{2}$ unchanged (relative to the situation without
interaction). Therefore, if the eigenstate-eigenvalue link is deemed necessary
in order to ascribe a value to a quantum system, then very clearly weak values
will not be able to ascribe quantum properties. Although to our knowledge, the
status of weak measurements has not been up to now explicitly discussed in terms of
property ascription relying on the eigenstate-eigenvalue link, it seems to us
that much of the criticism raised against weak values is implicitly relying on
this point.

For example for Leggett \cite{leggett} a weak measurement does not qualify as
\textquotedblleft\emph{a true measurement process}\textquotedblright, true
meaning here that the pointer states should be orthogonalized, hence leading to the
standard measurement described by Eq. (\ref{vn}). Sokolovski \cite{soko} requests that
measurements should create real pathways (calling for orthogonal pointer
states correlated with orthogonal eigenstates) as opposed to virtual pathways
(that take place when the system states in the pointer basis are not
orthogonal, leaving the property undefined). Svensson concludes his analysis
\cite{svensson} by asserting that weak values cannot represent
\textquotedblleft\emph{ordinary properties}\textquotedblright, on par with
eigenvalues. While Svensson does not discuss property ascription in quantum
mechanics nor mentions the eigenstate-eigenvalue link, it turns out (for reasons that will become clear in Secs. \ref{wvwv} and \ref{meaning}) that his
requirement of \ \textquotedblleft bona fide\textquotedblright\ properties can
only be fulfilled when the system ends up in an eigenstate of the measured observable.

\subsection{Weak values as ensemble expectation values}

The most common way of introducing weak values is to state \ they represent
some sort of expectation value in pre and post-selected ensembles; a detailed
exposition of this approach is given in \cite{dressel2012,ipsen2015}. The first argument in favor of this thesis is that
experimentally, a weak value can only be determined by measuring an ensemble
of identically prepared and post-selected systems.\ The shift is indeed very
small and can therefore not be meaningfully measured for a single system; the
weak value appears statistically as the average taken over the
ensemble.\ Second, as we have seen above, Eqs. (\ref{rea})-(\ref{ima}), the
real and imaginary parts of a weak value are formally equal to
conditional\ expectation values of different operators. Third, it can be shown
\cite{dressel2015} that the weak values define an operator that is the best
estimate of an observable $A$ when not only the initial state but the final
state is known\footnote{The estimate minimizes a specific distance $d$ in
Hilbert space, namely $d=\mathrm{Tr}\left[  \left\vert \psi(t_{i}%
)\right\rangle \left\langle \psi(t_{i})\right\vert \left(  A-A_{\mathrm{est}%
}\right)  ^{2}\right]  ,$ and the resulting best estimate is \cite{dressel2015} $A_{\mathrm{est}%
}=\sum_{f}\operatorname{Re}A_{f}^{w}\left\vert b_{f}\right\rangle \left\langle
b_{f}\right\vert \operatorname{Re}A_{f}^{w}$ where $A_{f}^{w}$ is the weak
value (\ref{wvt}).}.

In our view, none of these reasons are compelling. The first point appears as
a practical issue in which statistics are employed to reduce the measurement
uncertainties, and has no bearing on fundamental aspects.

The second argument relies on a numerical equivalence: the value of the shift,
given by $\operatorname{Re}A^{w}$, is equal to a conditional expectation
value, but this does not imply that $\operatorname{Re}A^{w}$ is itself an
expectation value, i.e. a statistical quantity relevant to ensembles. This can
be seen very easily in the particular case in which the pre and post-selected
states are arbitrary but identical. Then
\begin{equation}
A^{w}=\left\langle \psi(t_{w})\right\vert A\left\vert \psi(t_{w})\right\rangle
,
\end{equation}
so the weak value is numerically given by the expectation value. In this case
the pointer state (\ref{finalps}) is given by
\begin{equation}
\left\vert \varphi(t_{f})\right\rangle =\sum_{k}p_{k}(a_{k})e^{-iga_{k}%
P}\left\vert \varphi(t_{i})\right\rangle , \label{zz1}%
\end{equation}
with $p_{k}(a_{k})\equiv\left\vert \left\langle a_{k}\right\vert \left.
\psi(t_{w})\right\rangle \right\vert ^{2}$ as above. When $g$ is small it is
easy to see that the weighted superposition (\ref{zz1}) over the shifted
pointer states $ga_{k}$ results in the shift $g\left\langle \psi
(t_{w})\right\vert A\left\vert \psi(t_{w})\right\rangle $. This is the shift of a single pointer, obtained in a single run. 

The third point is an interesting observation, but depends on the choice of a
specific distance in Hilbert space (arguments based on the choice of a
different distance have been put forward to show the opposite, namely that
weak values do not behave as averages, see Sec. \ref{gev} below). Moreover, it
is difficult to explain how a\ physical pointer can be shifted by an optimal
estimator, which is by definition an epistemic quantity.

Therefore, leaving aside commitments to a fully epistemic interpretation of
the quantum formalism, for instance if one adheres to the statistical
interpretation of quantum mechanics \cite{ballentine} (by which it is assumed
that the quantum formalism intrinsically describes ensembles) there is no
ground to assert that weak values only characterize ensembles with post-selection.

\subsection{Weak values as generalized eigenvalues in a single
system\label{gev}}

Weak values were originally introduced \cite{AAV} as a generalized form of
eigenvalues, or rather \textquotedblleft\emph{a new kind of value for a
quantum variable}\textquotedblright\ \cite{AAV}. In our context, we will take
this to mean that (i) a weak value is a quantity relevant to a single system
(as opposed to an ensemble property); and (ii) a weak value is relevant to a
property of the quantum system, namely it gives the value of a quantum
observable correlated with a given post-selection. There are several arguments
in favor of this thesis.\ First, the pointer motion that is generated by the
weak value, see Eq. (\ref{finre}), is taken to be analogous to the pointer
motion proportional to an eigenvalue in the case of a projective
measurement.\ Second, the expressions (\ref{sav}) and (\ref{wav2}) give the
same observable average in terms of eigenvalues and weak values respectively;
in the latter case, the probability $p_{k}(b_{k})$ appearing in Eq.
(\ref{wav2}) is the probability of post-selecting to state $\left\vert
b_{k}\right\rangle $ assuming the disturbance induced by the weak interaction
can be neglected. An additional argument, that can be seen as a consequence of
the first, was recently given by Vaidman and co-workers \cite{vaidman-beyond}:
they examine the effect on the pointer dynamics when the shift is induced by
an eigenvalue, a weak value, or an average (the pointer is then in a mixed
state) and find that for short times the pointer with a weak value shift
behaves much more like an eigenvalue shifted pointer than the mixed pointer
state corresponding to an average value.

It is not difficult, if one agrees that an eigenvalue is a property of a
single system, to admit point (i) above. Indeed, upon post-selection the observable  $B$ undergoes a standard projective measurement and the
corresponding pointer at first entangled with the system ends up indicating the eigenvalue $b_f$ which we have assumed to be a property of a single system. Since the weakly coupled pointer is entangled with the system, which in turn becomes entangled with the post-selection pointer, the weakly coupled pointer undergoes a small shift upon post-selection. This shift must also be the property of a single system, since there is no reason to interpret the entanglement involving the weakly coupled pointer differently than the entanglement involving the post-selection pointer. In other words, this shift is an ``element of reality'', and hence the \textquotedblleft\emph{mechanical effect}%
\textquotedblright\ \cite{aharonov-botero} of the system on the weakly coupled pointer is therefore
established as being relevant to a single overall system. \ 

Whether this mechanical effect indicates a generalized eigenvalue
representative of a system property is not so straightforward. In the
\ specific case in which $A^{w}$ is indeed an eigenvalue -- implying that
either the pre-selected or the post-selected state is an eigenstate of the
weakly measured observable $A$ -- one relies indirectly on the
eigenstate-eigenvalue link: the eigenstate is either the pre or post-selected
state, and the eigenvalue comes out of the weak measurement by
orthogonalization (the pointer states are indeed orthogonalized despite their overlap).

In the general case, when both the pre and post-selected states are arbitrary,
the real and imaginary parts of $A^{w}$, given by Eqs. (\ref{rea}) and
(\ref{ima}) involve the ratio of averages because as we have seen, due to the
weakness of the interaction, the pointer captures the entire spectrum of the
weakly measured observable $A$. Moreover the expression does not involve the
sole weakly measured observable $A$, but the projector to the post-selected
state $\Pi_{b_{f}}$. Finally, we will argue below (Sec.
\ref{wvwv} and  \ref{meaning}) that the system has no element of reality  corresponding to $A^{w}$. For these reasons the term \textquotedblleft generalized
eigenvalue\textquotedblright\ might not be very appropriate to characterize a
weak value.

\subsection{Weak values as perturbation amplitudes}

The formal definition of the weak value [Eq. (\ref{wvt})] is given by a ratio
of amplitudes. This point has often been been put forward
\cite{kastner-sphmp,svensson,sokoB} in order to assert that weak values cannot
have any meaningful relevance to physical \ properties. We have already stated that any approach that relies, albeit
implicitly, on the eigenstate-eigenvalue link in order to ascribe properties
to a quantum system will consistently deny that amplitudes, and hence weak values, can represent
values of quantum properties.

Sokolovski goes further \cite{sokoB} in arguing that amplitudes are ubiquitous when perturbation theory is
applied, and sees weak measurements are a specific output of perturbation
theory. This is of course indisputable from a technical point
of view, but such arguments do not take into account the peculiar character of this form of perturbation theory, that is almost identical to a standard measurement process and  induces pointer shifts. In this sense, this type of criticism appears as incomplete \cite {cohen}.

\subsection{Weak values as weak values\label{wvwv}}

A standard projective measurement of a property represented by the observable
$A$, of the type described above (see Sec. \ref{eel}), involves a correlation
between the pointer position and an eigenvalue of $A$. The entangled state
(\ref{vn}) between the pointer and the system correlates each eigenvalue with a
an unambiguously discriminate pointer state.\ At the end of the measurement
process (after the projective collapse), the pointer indicates the value of
the system property.

In a weak measurement, the weakly coupled pointer similarly indicates $\operatorname{Re}A^{w}$, but the shift is small and appears after the post-selection collapse, whereby the post-selection
pointer indicates the value $b_{f}$ of the property corresponding to the
observable $B$. Hence unlike an eigenvalue, $\operatorname{Re}A^{w}$ does not reflect the
value of the sole property $A$, but the value of $A$ correlated with the
system having the eigenvalue $b_{f}$ for the property
$B$. Moreover, although $\operatorname{Re}A^{w}$ depends on the time $t_{w}$
and on the location of the interaction zone with the weakly coupled pointer,
the weak value only appears at the post-selection time $t_{f}$.\ But at
$t_{f}$ the system has a value $b_{f}$ for the property $B$ and no value can
be ascribed to the property $A$. Strictly speaking the (real part of the) weak
value does not ascribe a property to the system, in the sense that there is no
corresponding element of reality in the system.

Nevertheless the state of the weakly coupled pointer upon post-selection can
be predicted with certainty and is an element of reality for the pointer.
This results from a mechanical effect of the coupling interaction on the pointer, that we derived in Sec. \ref{protocol} and that can also be shown to follow from the dynamics of the pointer variable in the Heisenberg
picture \cite{aharonov-botero}. This mechanical effect  
characterizes the value of $A$ when the system is filtered to an eigenstate of a different observable.

This is exactly how the expression giving
$\operatorname{Re}A^{w}$ [Eq. (\ref{rea})] can be read: the relevant observable is $\frac{1}%
{2}\left(  \Pi_{b_{f}(t_{w})}A+A\Pi_{b_{f}(t_{w})}\right)$, a symmetrized operator describing the measurement of $A$ followed by
$\Pi_{b_{f}}$. A well-known quantity employed in standard quantum mechanics that has this form is the Schr\"odinger current $j_\psi(x,t)$, with the corresponding operator being given by \cite{cohendiulaloe} $J=\frac{1}{2m}\left( \left\vert x\right\rangle \left\langle x\right\vert P+P\left\vert x\right\rangle \left\langle x\right\vert \right) $, where $P$ is the momentum operator \footnote{Unsurprisingly, the current density appears in the numerator of the following weak value of the momentum, $\mathrm{Re}\frac{\left\langle x\right\vert P\left\vert \psi (t)\right\rangle }{\left\langle x\right\vert \left. \psi (t)\right\rangle }=\frac{mj(x,t)}{\rho (x,t)}$.}. 
The denominator in  Eq. (\ref{rea}) accounts for the renormalization of the density $\rho=\left\vert \psi(t_{w})\right\rangle
\left\langle \psi(t_{w})\right\vert $, as only the fraction of $\rho$
that reaches post-selection is to be taken into account. \\   Note that in a purely classical context, the
expression equivalent to Eq. (\ref{rea}) would represent \cite{matzkin-prep}
the motion of a pointer coupled to the system through the classical
interaction Hamiltonian (\ref{Hint}), when a filter is implemented. This
filter selects the classical particles that will have a specific value $b_{f}$
at some final time $t_{f}$, after the weak interaction\footnote{The
corresponding classical expression is
\begin{equation}
\int_{\mathcal{B}_{f}}A(q,t_{w})\frac{\rho(q,t_{w})}{\int_{\mathcal{B}_{f}%
}\rho(q^{\prime},t_{w})dq^{\prime}}dq
\end{equation}
where $\rho(q)$ is the configuration space classical distribution. The
integral is taken over $\mathcal{B}_{f}$ which is the set of all $q$'s taken
at $t_{w}$ such that at the final time $t_{f}$ we have $B(q,t_{f})=b_{f}$.\ In
a classical setting, the filtering needs to be done before the weak
interaction takes place.\ Note that the denominator is simply the
normalization constant for the density due to the filtering (see
\cite{matzkin-prep} for details).}. Quantum mechanically the filter is the post-selection, and
the apparent retrodictive aspect arises upon post-selection from the quantum
correlations imprinted in the entangled state (\ref{finalps}) between the
system and the pointer.

\subsection{The meaning of weak values\label{meaning}}

\ We have argued that, despite similarities with eigenvalues, property
ascription for weak values is not straightforward. Indeed, the state of the
weakly coupled pointer after post-selection (at time $t_{f}$) can be predicted
with certainty -- it is an element of reality as per Sec. \ref{er} -- but
regarding the system only the post-selected state $\left\vert b_{f}%
\right\rangle $ is an element of reality. Hence for the system there is no
element of reality corresponding to the weak value, neither at the time
$t_{w}$ of the interaction, nor at post-selection.\ This is hardly surprising
since the system state is minimally disturbed by the interaction at $t_{w}$
and has acquired the property value $b_{f}$ after post-selection.

Despite the lack of an element of reality in the system corresponding to a
weak value, it remains possible to link the shift $\operatorname{Re}A^{w}$ to
a form of system property.\ As we have seen in Sec. \ref{wvwv}, this link is
embodied in the correlations encapsulated in the entangled system-pointer
state (\ref{finalps}). The weak value -- that is the mechanical effect on the
weakly coupled pointer described in Sec. \ref{gev} -- reflects retrodictively
the value of $A$ due to the coupling (that took place at the earlier time $t_{w})$
compatible with post-selection. In this sense it is a partial property of the
system, relative to a specific space-time region (defined by the location of
the weakly coupled pointer) and relative to a choice of post-selected
observable and eigenvalue. This form of property ascription is considerably
weaker than the one for eigenvalues, which holds for the entire system and is
grounded on the existence of a corresponding element of reality.

Nevertheless, this weaker form of property ascription can be meaningful and
useful. We have mentioned above the Schr\"odinger current density as a well known quantity in standard quantum mechanics having the same structure as weak values. It can hardly be maintained that the current density at a particular space-time point does not characterize a partial property of the system at that particular space-time point.  We have amply discussed elsewhere \cite{duprey2017,duprey2018} the
case of null weak values. In the case of a projector $A=\left\vert
a_{k}\right\rangle \left\langle a_{k}\right\vert ,$ a null weak value
$A^{w}=0$ means that the property represented by $A$ cannot be registered by
the weakly coupled pointer for the given post-selection. Such a result stems
from the quantum correlations between the weakly coupled and post-selection
pointers, and also holds for a strongly coupled intermediate pointer. Null
weak values have been used to interpret phenomena like the Quantum Cheshire
Cat \cite{aharonov-qcc,matzkin-pan} or to account for discontinuous
trajectories in the proposals investigating the past of a quantum particle
\cite{vaidman-past,exp-pqp}.

Anomalous weak values (that is WV falling outside the eigenvalue range, such
as $A^{w}=-1$ for a projector) are a consequence of the non-commutativity of
the projectors into the pre-selected and post-selected states and the
observable $A$. They are intimately linked to interference effects and cannot
be obtained with classical probability distributions \cite{dressel2015}. As we have
argued in this paper, the interpretation of anomalous weak values as
\emph{bone fide} properties on par with eigenvalues \cite{svensson} cannot
hold: there is no corresponding element of reality in the system, as a weak
value describes a partial property at a given space-time point,
characterizing amplitudes and depending on interference effects. Anomalous weak
values still have explanatory power when the system is considered as a
whole.\ For instance a negative projector value or a negative particle number
on a given path may not be particularly illuminating by itself, but comparing
with weak values of the analogous projectors on other paths gives an
explanation -- in terms of experimentally measurable
quantities -- of the dynamics of interference, and further explains the outcomes obtained when projective
measurements are made at an intermediate time. Last but not least, weak values
give an additional experimentally observable confirmation of the validity of
the standard quantum formalism at the level of transition amplitudes, as
measured by weakly coupled pointers.

\section{Conclusion \label{concl}}

We have investigated in this work the meaning of weak values through the prism
of the description of the properties of a quantum system that evolves from an
initially prepared state to a final post-selected one. We first recalled how
properties are ascribed to quantum systems, namely through the
eigenstate-eigenvalue link. We focused on pre and post-selected systems to
examine how the eigenstate-eigenvalue links works when attempting to
understand the property of a quantum system at an intermediate time. The
emerging picture is somewhat limited, since such intermediate properties
depend on the measurements that are made, while any attempt to unify the
physical picture by  counterfactual reasoning leads to paradoxes.

The weak measurements framework bypasses these limitations by implementing a
minimally perturbing interaction with a quantum pointer. The weak value,
quantifying the imprint of the interaction and the subsequent post-selection
on the pointer, shares some similarities with eigenvalues, in particular the
fact that, if an eigenvalue is assumed to be relevant to a single system (and
not an ensemble), then this is also the case for a weak value. We examined
property ascription to a system observable based on weak values.\ This turned
out to be a subtle issue, as a weak value $A^{w}$ characterizes the system
observable $A$ filtered by post-selection in a retrodictive manner, mediated
by entanglement and without a corresponding system element of reality.

We discussed several interpretations that have been given to weak values, and
argued that weak values can indeed be seen as ascribing properties to a system
but in a partial way, certainly not on par with the standard property
ascription based on the eigenstate-eigenvalue link. The explanatory power
afforded by the weak measurements framework not only concerns the outcomes
obtained in standard projective measurements when quantum interferences play a
prominent role, but confirm the validity of the standard formalism at the
level of amplitudes.\ In turn, this could lead to novel fascinating implications
concerning the physical nature of the formalism described by quantum theory.

\textbf{Acknowledgment}:Dipankar Home (Bose Institute,Kolkata) and Urbasi Sinha (Raman Research Institute, Bangalore) are thanked for useful discussions on an earlier version of the manuscript. Partial support from the Templeton Foundation (Project
57758) is gratefully acknowledged.

\end{document}